\documentclass[namedreferences]{solarphysics}
\usepackage[natbib,optionalrh,solaromanenum]{spr-sola-addons} % For Solar Physics 
\usepackage{graphicx}                    % For eps figures, newer & more powerfull
\usepackage{amssymb}                    % useful mathematical symbols
\usepackage{color}                       % For color text: \color command
\usepackage{url}                         % For breaking URLs easily trough lines
\newcommand{\pref}{\protect\ref}
\newcommand{\ddx}[2]{\frac {\partial #2}{\partial #1} }

\begin{document}

\begin{article}

\begin{opening}

\title{From Forbidden Coronal Lines to Meaningful Coronal Magnetic Fields}

%%%%%%%%%%%%%%%%%%%%%%%%%%%%%%%%%%%%%%%%%%%%%%%%%%%
%% Authors Names
%
\author{P.G.~\surname{Judge}$^{1}$\sep
        S.~\surname{Habbal}$^{2}$\sep
        E.~\surname{Landi}$^{3}$      
       }

%%%%%%%%%%%%%%%%%%%%%%%%%%%%%%%%%%%%%%%%%%%%%%%%%%%
%% Runningheads
%
\runningauthor{P.G.~Judge {\em et al.}}
\runningtitle{Coronal Magnetic Fields}

%%%%%%%%%%%%%%%%%%%%%%%%%%%%%%%%%%%%%%%%%%%%%%%%%%%
%% Affilations 
%
  \institute{$^{1}$ High Altitude Observatory,  P.O. Box 3000,
    Boulder, CO 80307-3000, USA
                     email: \url{judge@ucar.edu} \\ 
             $^{2}$ Institute for Astronomy,
2680 Woodlawn Drive, Honolulu, HI 96822, USA
                     email: \url{shadia@ifa.hawaii.edu} \\
             $^{3}$ Department of Atmospheric, Oceanic and Space
             Sciences, University of Michigan, Ann Arbor, MI 48109, USA
                     email: \url{elandi@umich.edu} \\
             }

%%%%%%%%%%%%%%%%%%%%%%%%%%%%%%%%%%%%%%%%%%%%%%%%%%%
%%% Abstract 
\begin{abstract}
We review methods to measure magnetic fields within the
corona using the polarized light in magnetic-dipole (M1) lines.  We are
particularly  interested in both the global magnetic-field evolution over a
solar cycle, and the local storage of magnetic free energy within 
coronal plasmas.  We address commonly held skepticisms concerning
angular ambiguities and line-of-sight confusion. We argue that 
ambiguities are in principle no worse than more familiar remotely
sensed photospheric vector-fields, and that 
the
diagnosis of M1 line data would benefit from simultaneous 
observations of EUV lines.  Based on calculations and data from eclipses, we
discuss the most promising lines and different approaches
that might be used.  We point to the S-like [Fe~{\sc XI}] line (J=2 to J=1) at
789.2nm as a prime target line (for ATST for example) to augment the
hotter 1074.7 and 1079.8 nm Si-like lines of [Fe~{\sc XIII}] currently
observed by the Coronal Multi-channel Polarimeter (CoMP).  Significant
breakthroughs will be made possible with the new generation of
coronagraphs, in three distinct ways: (i) through single point
inversions (which encompasses also the analysis of MHD wave modes),
(ii) using direct comparisons of synthetic MHD or force-free models
with polarization data, and (iii) using tomographic techniques.
\end{abstract}

%%%%%%%%%%%%%%%%%%%%%%%%%%%%%%%%%%%%%%%%%%%%%%%%%%%
%% Keywords
%
%\keywords{}

\end{opening}
%-------------------------------------------------

\newcommand\pmag{\ifmmode \omega_{\rm L}/\Delta\omega_{\rm D}
\else$\omega_{\rm L}/\Delta\omega_{\rm D}$\fi}
\renewcommand{\v}[1]{\ifmmode {1\over2}(3\cos^2#1-1)\else ${1\over2}(3\cos^2#1-1)$\fi}
\newcommand{\tb}{\ifmmode \Theta_{\rm B}\else $\Theta_{\rm B}$\fi}
\newcommand{\gb}{\ifmmode \gamma_{\rm B}\else $\gamma_{\rm B}$\fi}
\newcommand{\vtb}{\ifmmode \vartheta_{\rm B}\else $\vartheta_{\rm B}$\fi}
\newcommand{\vpb}{\ifmmode \varphi_{\rm B}\else $\varphi_{\rm B}$\fi}
\newcommand{\vtm}{\ifmmode \vartheta_{\rm M}\else $\vartheta_{\rm M}$\fi}
\newcommand{\pop}[1]{\ifmmode N(\alpha_0 {\scriptstyle #1}) \else $N(\alpha_0 {\scriptstyle #1})$\fi}
\newcommand\ft{l(\vtm)}
\newcommand{\vk}{\hat{\bf k}}

\newcommand\ec{\varepsilon}
\newcommand\iec{\Phi}

\newcommand\alignment{\sigma^2_0(\alpha_0 J)}

\newcommand\alig{k_J(T_{\rm e},n_{\rm e},\vtm)\,\v{\vtb} \,}
\newcommand\modalig{k_J(T_{\rm e},n_{\rm e},\vtm)|\v{\vtb}| \,}

\newcommand\freq{\nu}
\newcommand\frt{({\bf r};t)}
\newcommand\brt{{\bf B}({\bf r};t)}

%%%%%%%%%%%%%%%%%%%%%%%%%%%%%%%%%%%%%%%%%%%%%%%%%%%
%% Sections
%
\section{Introduction}\label{s:Introduction} 

Measurement of solar magnetic-fields has been a goal of solar physics
since the discovery of the Zeeman effect in sunspots by 
\cite{Hale1908b}.  Our purpose here is to review how magnetic-dipole (M1) lines,
formed in coronal plasma, 
might be used to address particular questions in coronal and
heliospheric physics: How does the coronal magnetic-field vector
evolve over the solar sunspot cycle?  Can we measure some of the free
magnetic energy on observable scales in the corona, and its changes,
say, before and after a flare?  

Theoretical work by  \cite{Charvin1965} spurred experimental studies of
the polarization of magnetic-dipole lines, such as [Fe~{\sc XIII}]
$3p^2\ ^3\!P_1 \rightarrow 3p^2 \ ^3\!P_0$ at 1074.7\,nm, as a way to
constrain coronal magnetic-fields.  The lines are optically thin in
the corona; their intensities are $\lesssim 10^{-5}$ of the disk
continuum intensities.  Thus they can be observed only during eclipses
or using coronagraphs that occult the solar disk.

Here, we review M1 emission-line polarization towards 
the specific goal of
{\em measuring the vector magnetic-field
  $[\brt]$ throughout a sub-volume of the corona}. 
To date, this has not been
achieved.  We have little idea of the true origin of CMEs, flares,
and coronal heating. even though coronal plasma has been regularly observed since the
1930s. The latest of several decades of high-cadence images of coronal plasma from space reveal
more details but are limited to studying effects, not causes, of
coronal dynamics, since such instruments measure thermal, not magnetic,
properties.  To discover the {\em cause} of coronal dynamics 
we {\em must  measure} ${\bf B}\frt$ {\em above the photosphere- the region of the
  atmosphere where free energy is stored and quickly released}, since it is the free energy
associated with electrical current systems within coronal plasmas that 
drives these phenomena.   Measurements of ${\bf B}\frt$ in the
photosphere have been done for decades, but photospheric dynamics
occurs under mixed $\beta$ conditions ($\beta=$ gas/magnetic pressure
$\approx 1$).   In
contrast, the low-$\beta$ coronal plasma should exist in simpler
magnetic configurations, perhaps more amenable to straightforward
interpretation. 
In MHD the electrical currents are simply
${\bf j} = {\rm curl}~{\bf B}\frt$.  Given sufficiently accurate 
measurements of ${\bf B} \frt$ in the low-$\beta$ corona, both
${\bf j}$ and the free energy
itself can in principle be derived.  

Like all observational studies, this is bandwidth-limited exercise.
We can investigate structures only from the smallest resolvable scales
$\ell$ to the largest $\approx {\rm R}_\odot\approx 700$ Mm, and on time scales longer than
the smallest time $\tau$ needed to acquire the data. The
spatial range will be limited by forseeable observational capabilities
to $ \ell \gtrsim 1$Mm.  Successful tomographic-inversions using 
solar rotation to slice through the 3D corona require $\tau \gtrsim 1$
day, during which the corona is viewed from angles differing by $\approx1/4$
radian.  Given our goal, it is clear that we will not be able to
investigate either the dissipation scales of magnetic-fields, nor
changes in magnetic-fields on rapid dynamical time scales $\lesssim
{\rm R}_\odot/C_A \approx 350$\,s of the inner corona (here $C_A \approx 2$ Mm s$^{-1}$ is
the Alfv\'en speed).
However, these limitations are not new.  In any case coronal
dynamics and flares involve a slow build-up and sudden
release of magnetic free energy \citep{Gold+Hoyle1960}.  This  energy
build-up can indeed, and should be, explored through
new measurements of ${\bf B} \frt$.

\section{The Inverse Problem}
\label{sec:inversion}

\subsection{General Considerations}

Consider a heliocentric coordinate system with
Sun center at ${\bf r}\equiv(x,y,z)^T={\bf 0}$, with the line of sight along
$z$, and $x,y$ being in the plane of the sky.  Given a set of observations
$\left [\{I_{i\ldots4,\nu}(x,y,t)\}\right ]$ of the four Stokes parameters [$IQUV$] at $n$
frequencies $[\nu]$ across a M1 line
at time $t$, we seek solutions for ${\bf
  B}(x,y,z;t) $ over an observable sub-volume of the corona $\approx \Delta x
\Delta y \Delta z$.  We can write
\begin{equation} \label{eq:forward}
I_{i,\nu}(x,y;t) = \int_{\Delta z} \ec_i({\bf S}\frt)\, {\rm d}z  =
\iec_i({\bf S}\frt).
\end{equation}
\noindent The M1 lines -- having small oscillator strengths -- 
are optically thin through the corona.  Under these conditions 
$\ec_i$ is a non-linear, but 
{\em local} function of a ``source vector'' $S_j,j=1\ldots n$.  
The price for ``optical thinness'' is that $\Delta z$ encompasses 
the entire line of sight through the corona to the solar disk or into
space.  Tomography specifically takes advantage of
this.  There is, however,  skepticism in the community 
concerning the magnetic-field measurements under optically thin
conditions 
that we 
address in Section \pref{s:LOS}.  
The non-linearity arises because the Stokes parameters ${\bf I}$ 
depend on the ``atomic alignment'' [$\alignment$], a scalar 
quantity that is a linear combination of magnetic-substate
populations. The alignment can be positive, negative, or zero, as discussed below.  
Ignoring the alignment would make the problem linear in the source 
term (like the standard emission measure problem for line
intensities only).

The source [${\bf S}$] must be written as a function of ${\bf r}$ and
time [$t$] in terms of necessary thermodynamic and magnetic
parameters. At a minimum this means specifying $${\bf S \frt}  = \left
  \{\rho\frt,{\bf
  v\frt},
T\frt,{\bf B}\frt \right \},$$ for
plasma with density $\rho$ moving with velocity ${\bf v}$
at temperature $T$.   These quantities must be supplemented by
calculations that give the local distributions of ionization states
and electron density.  
Any formal ``inverse'' solution is
of the form
\begin{equation} \label{eq:inverse}
{\bf S}\frt = \iec^{-1} {\bf I}(x,y;t),
\end{equation}
 where ${\bf I}$ is the $4n$-long ``vector'' of observed Stokes
 parameters.   Clearly, a 3D array of scalar and vector-fields 
such as ${\bf S}\frt$ cannot be recovered from one
set of measurements [${\bf I}(x,y;t)$] that are integrated over 
$\Delta z$.  Additional information is needed.

A ``good diagnostic'' maps components of ${\bf I}$ into ${\bf S}$. If
Equation~(\pref{eq:forward}) were linear (or were linearized) 
we could write 
\citep[{\em e.g.,\rm}][]{Craig+Brown1986}
\begin{equation} \label{eq:cb}
{\bf S} = {\bf \left (\iec^T\iec \right )}^{-1} {\bf \iec^T \ I}.
\end{equation}
The eigen-spectrum of matrix ${\bf \left (\iec^T\iec \right )}^{-1} $ 
measures the degree to which measurements of ${\bf I}$ 
can be used to determine ${\bf S}$.   As usual, the formal operation
given by Equation (\pref{eq:cb}) should not be taken as an inverse
solution, it is ill-posed \citep{Craig+Brown1986}.  

\subsection{Origin of Polarization of Magnetic-Dipole Coronal Lines}
\label{s:info} 

Polarization of spectral lines is generated in two ways ({\em e.g.},
\citealp{Casini+Landi2008}).  Any process that produces unequal sub-level
populations, such as anisotropy of illuminating radiation, also
produces polarization of light in the emitted radiative transitions
to/from a given atomic level.  When magnetic-substate populations are
equal, the state is ``naturally populated'' and light is unpolarized.
The second way is to separate the substates in energy, so that
spectroscopy can discriminate states of polarized light
associated with the specific changes in energy of states with
different sub-level quantum numbers $[M]$, no matter how the sub-levels are 
populated. 
Magnetic- and electric- fields thus are imprinted on spectral line
polarization through the Zeeman and Stark effects.  Since charge
neutrality is a good approximation in coronal plasma
({\em e.g.} \citealp{Parker2007}), 
electric-fields and
the associated stresses are far smaller than those for the magnetic-field, and in quasi-static situations can be ignored. We focus on the
magnetic-fields.

Adopting the notation of
\cite{Casini+Judge1999}, for 
M1 emission-lines between upper- and lower-levels with quantum numbers $\alpha J$
($J=$ total angular momentum) and $\alpha_0 J_0$, 
the $\ec_i$ terms 
in Equation (\pref{eq:forward}) are proportional to a term
of the form 
\begin{equation}
\epsilon_{ J J_0} ={{h \freq}\over 4\pi}\,N_{\alpha_0
  J}\, A_{\alpha_0 J\to \alpha_0 J_0}. \end{equation}
This term is simply the emission
coefficient (ignoring stimulated emission) for the {\em unpolarized}
transfer problem, in units of erg~cm$^{-3}$~sr$^{-1}$~s$^{-1}$.  
The population density of the upper-level 
can be factored as usual as 
\newcommand{\gap}{\,}
\begin{eqnarray}
  N_{\alpha_0  J}\, & =&  \frac{N_{\alpha_0 J}\,}{N_{ion}} \gap
  \frac{N_{ion}}{N_{el}}  \gap \frac{N_{el}}{N_H}  \gap 
  \frac{N_H}{n_{\rm e}}  \gap  n_{\rm e} \gap.\\
\end{eqnarray}
We label the first factor on the RHS of the above equation $f$, it is 
the ratio of the upper-level population of the level
emitting the photons to the total ion population.  
The remaining factors are, in order, the ionization fraction, element abundance,
ratio of hydrogen nuclei number density to the electron number density
$n_{\rm e}$, and lastly $n_{\rm e}$ itself.  
For strong lines
(electric dipole or ``E1'' lines in the 
EUV/ soft X rays) $f \propto n_{\rm e}$ and 
$\exp(-h\nu/kT)$, so that $  N_{\alpha_0  J} \propto n_{\rm e}^2 G(T)$ as usual.  
For M1 lines $f \propto n_{\rm e}^\beta$ with $0<\beta
< 1$, but generally $h\nu/kT \ll1$,  so the temperature
dependence of $N_{\alpha_0  J} $ enters mostly the ionization fraction.  
Under coronal ionization equilibrium conditions this factor 
is a function only of temperature $T$.

The polarized terms [$\ec_i$]
also depend on the anisotropy of the incident
photospheric radiation, particle collisions, the strength and
direction of the coronal magnetic-field, and the direction of the line
of sight.  M1 lines have large
radiative lifetimes ($\tau_R \approx A_{\alpha_0 J\to \alpha_0 J_0}^{-1}
\approx 10^{-1}$s).  The Larmor frequency  
[$\nu_{\rm L} \approx \mu_B B/h$]
is much larger than the inverse lifetime of the level, 
$\nu_{\rm L} \tau_R \gg 1$.  This is the ``strong
field'' (or ``saturation'') limit of the Hanle effect.  
If the photospheric irradiation is rotationally symmetric and spectrally
flat, 
the atomic polarization is in the special form of 
alignment [$\alignment$] which in terms of substate populations is written
\begin{eqnarray} \label{eqn:alignpop}
\sigma^2_0(\alpha J) &=&
{\sqrt{5}\over\sqrt{J(J+1)(2J-1)(2J+3)}}
\sum_M [3M^2-J(J+1)]\,\frac{\pop{JM}}{\pop{J}}
\label{eqn:align2}
\end{eqnarray}
Circularly polarized light is generated only by the
``$\sigma$''-components ($\Delta M =\pm1$) of the Zeeman effect.
The M1 emission coefficients $\ec_i^{(j)}$
for Stokes parameter $i$ are (Section 4 of \citealp{Casini+Judge1999}):
\begin{eqnarray}
\ec_0^{(0)}(\freq,\hat {\bf k})
&=&\epsilon_{ J J_0}\,\phi(\freq_0-\freq)
	\left[1+D_{J J_0}\,\sigma^2_0(\alpha_0 J)\,
	{\cal T}^2_0(0,\hat{\bf k})\right]\;,
\label{eqn:eps00} \\
\noalign{\smallskip}
\ec_i^{(0)}(\freq,\hat {\bf k})
&=&\epsilon_{ J J_0}\,\phi(\freq_0-\freq)\,D_{J J_0}\,
	\sigma^2_0(\alpha_0 J)\,
	{\cal T}^2_0(i,\hat {\bf k})\;,
	\qquad (i=1,2)
\label{eqn:epsi0} \\
\noalign{\smallskip}
\ec_3^{(1)}(\freq,\hat {\bf k})
&=&-{\textstyle\sqrt{2\over 3}}\,\freq_{\rm L}\,
	\epsilon_{ J J_0}\,\phi'(\freq_0-\freq)
	\left[\bar{g}_{\alpha_0 J,\alpha_0 J_0}+E_{J J_0}\,
	\sigma^2_0(\alpha_0 J)\right]
	{\cal T}^1_0(3,\hat {\bf k}),
\label{eqn:eps31} 
\end{eqnarray}
\noindent 
where $j$  in  $\ec_i^{(j)}$ 
is the leading order in the Taylor series expansion of the
emission coefficient with frequency.  
\footnote{For
M1 coronal lines the
$\Delta M=0$ ``$\pi$''-components are proportional to 
$\phi''(\freq_0-\freq)$. These are orders of magnitude weaker than
the zeroth-order alignment-generated component, which is 
$\propto \phi(\freq_0-\freq)$.}  
$\phi(\freq_0-\freq)$ is the
line profile [Hz$^{-1}$], $\phi'(\freq_0-\freq)$ its first
derivative with respect to $\freq$,  remaining terms (except $\nu_{\rm
  L}$) 
are dimensionless. The factor
$D_{J J_0}$ depends only on angular momenta and $E_{J J_0}$ also
depends on the Land\'e g-factor of the transition: $\bar{g}_{\alpha_0
  J,\alpha_0 J_0}$.  The tensor ${\cal T}^{1,2}_0(i,\hat {\bf k})$
relates the angular distribution and 
polarization of emitted radiation to the direction
of the observer.  In terms of angles $\gamma_B$ and $\Theta_B$
defining the magnetic 
azimuth in the plane-of-the-sky and inclination along the
line-of-sight, these are 
\begin{center}
\begin{tabular}{lll}
${\cal T}^2_0(0,\vk)_{\rm M1}
$ &=&${\textstyle{1\over 2\sqrt{2}}}(3\cos^2\!\Theta_B-1)$\\
 ${\cal T}^2_0(1,\vk)_{\rm M1}
$ &=&${\textstyle{3\over 2\sqrt{2}}}\cos 2\gamma_B \sin^2\!\Theta_B$ \\
${\cal T}^2_0(2,\vk)_{\rm M1}
$&=&$-{\textstyle{3\over 2\sqrt{2}}}\sin 2\gamma_B \sin^2\!\Theta_B$\\
${\cal T}^1_0(3,\vk)_{\rm M1}
$&=&${\textstyle\sqrt{3\over 2}}\cos\Theta_B$.
\end{tabular}
\end{center}

\subsection{M1 Lines From One Point in the Corona}

Sometimes coronal images are dominated by emission from one small
region, 
such as from a small section of an active region loop at
${\bf r_0} = (x_0,y_0,z_0)^T$.  In this case the source ${\bf S}\frt\,
= {\bf S}(t) \delta(x-x_0) \delta(y-y_0) \delta(z-z_0)$ and the
measured {\bf I} is simply $\propto \ec_i^{(j)}$ evaluated at 
${\bf r} = (x_0,y_0,z_0)^T$.  By inspection of expressions for 
$\ec_i$  we see that
\begin{enumerate}
\item {} The magnetic-field {\em strength} is encoded only in circular 
polarization through $\ec_3^{(1)}(\freq,\hat {\bf k})$, via $\nu_{\rm L}$, and only as the
{\em product} $B\cos\Theta_B$.  
\item{}  The usual weak-field ``magnetograph
formula'' -- taking the ratio of Equation 
(\pref{eqn:eps31}) and the derivative of Equation (\pref{eqn:eps00}) -- 
does not only depend on the 
Land\'e g-factor $\bar{g}_{\alpha_0
  J,\alpha_0 J_0}$.  In the presence of a non-zero alignment, 
the ratio includes smaller terms including 
$\alignment$ in both numerator and denominator.  
\item {} The magnetic-field {\em azimuth} $\gamma_B$ is encoded in the linear 
polarization as $\gamma_B =$ $-\frac{1}{2} 
\arctan (\ec_2^0/\ec_1^0)$.
%\item{} The atomic alignment $\sigma$ is an important parameter and is
 % discussed next (section \pref{sec:align}).
\end{enumerate}
Of course, in reality, measured quantities [{\bf I}]
are integrals of these elementary
$\ec_i^{(j)}$ coefficients along the line of sight.   

\subsection{``Long'' Line of Sight Integrations}\label{s:LOS} 

A  concern sometimes expressed among solar physicists is that M1 coronal
emission-lines form over such large distances that they have limited
use in diagnosing magnetic\--fields.  The perceived problem is that the
magnetic-field changes too much along the long lines of sight $L_c
\approx {\rm R}_\odot$.  Mathematically we might say
\begin{equation}{\label{e:scales}}
\left |\ddx{s}{B_i}\right |L_c  \gtrsim |\langle B_i\rangle|,
%,  \ \ {\rm or} \ \ 
%\sigma(B_i) \gtrsim |\langle B_i\rangle|.
\end{equation}
for magnetic vector component $B_i$.

\begin{figure} 
 \centerline{
\includegraphics[width=0.5\textwidth,clip=0]{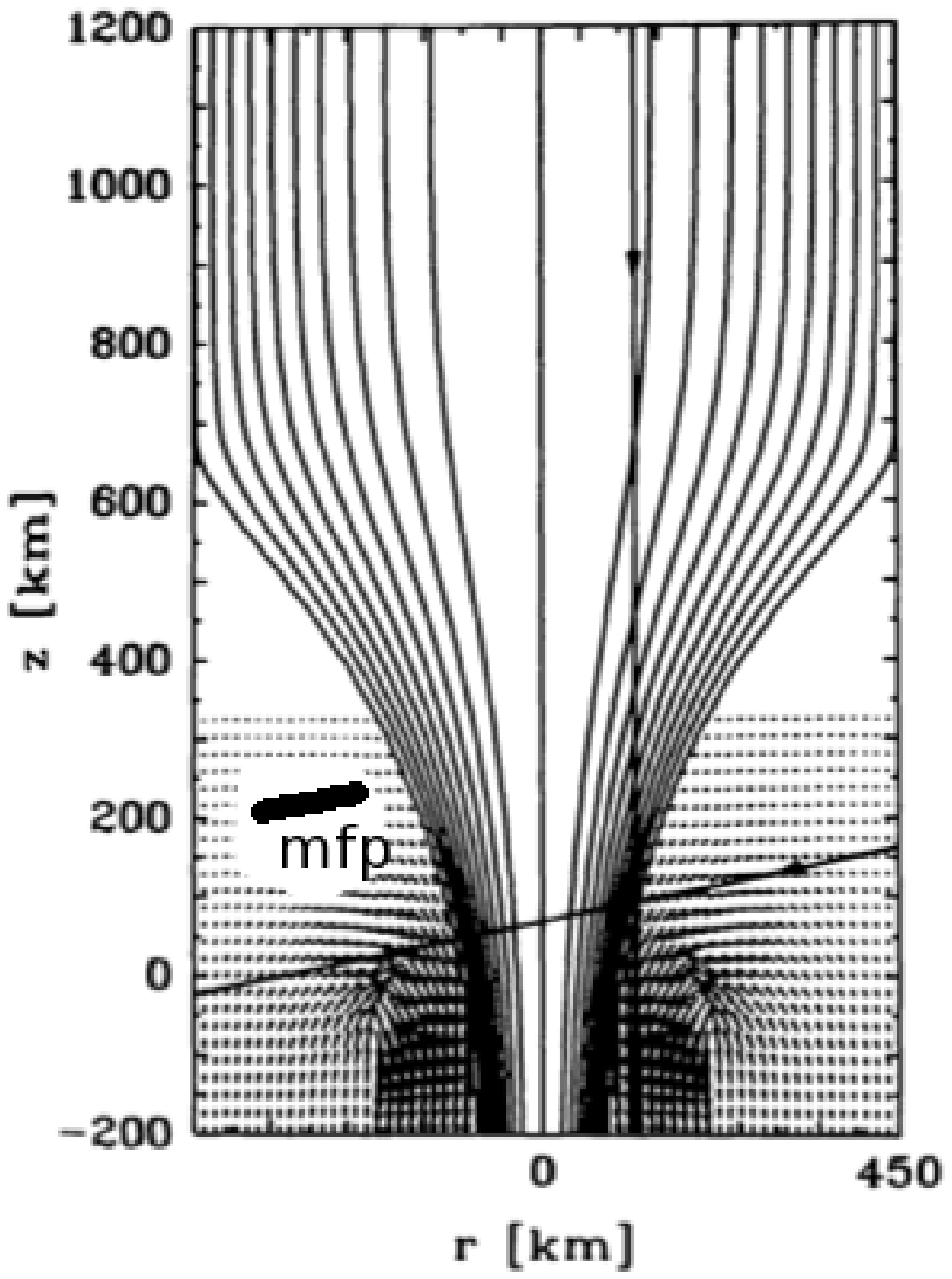}
\includegraphics[width=0.5\textwidth,clip=0]{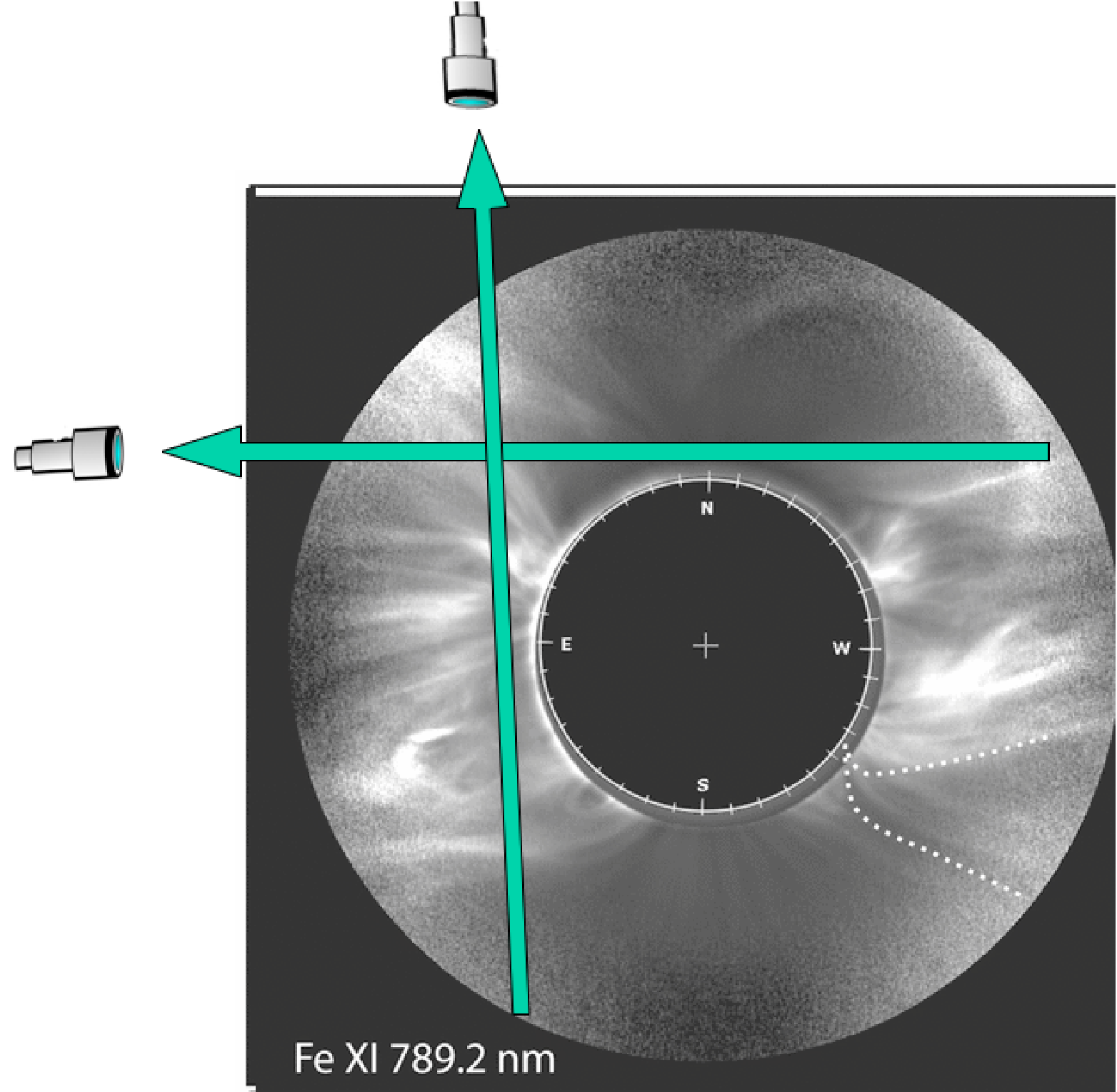}
}
 \caption{Left: magnetic-field lines and velocity vectors 
for a flux tube extending from beneath the photosphere  into the 
chromosphere, from Steiner (1994) but annotated with the photon mean
free path.  Two rays intercepting the
   boundary between magnetised plasma are shown, along which 
spectral lines are formed.  
Right: an image of the corona during the July 2011 eclipse
   in the Fe {\sc XI} 789.2\,nm line is shown, obtained by one of us (SH). 
Two integration rays are shown.  Note that this image
   already has been integrated in one dimension, thus in 3D such rays intercept
   much less structure than this image might seem to suggest.  
}\label{fig:ftcor}
\end{figure}

Let us apply the same arguments to a familiar situation in which
there is far less such preconceived skepticism: the solar 
photosphere.   It is indeed a ``thin'' layer (500\,km) compare with 
the solar radius, but 
this does not mean that it is  ``thin''  (small $L_{\rm c}$) 
in the sense implied by 
Equation~(\pref{e:scales}). 
Photospheric magnetic-fields are
highly intermittent in space and time.  Consider  
formation of polarized
light from a simple cylindrical ``flux tube'' of diameter
160\,km in the solar photosphere ({\em e.g.} \citealp{Steiner1994}, left panel
of Figure~\pref{fig:ftcor}).  The photon mean free path [mfp] in the
photosphere $L_p\approx H_p$ is $\approx 120$ km, as indicated by
``mfp''.
 Clearly there is structure in the
thermal and magnetic conditions {\em well below the photon mfp}.  As
discussed by Steiner and others, this leads to ``peculiar'' Stokes
profiles -- the ``Stokes-$V$ area asymmetry'' being one parameter 
of particular
interest.  The point here is not that peculiar Stokes profiles can be
explained, but that in photospheric problems of interest, one 
must diagnose magnetic-fields in situations the inequality in 
Equation~(\pref{e:scales}) holds!  

Consider next the second panel of Figure~\pref{fig:ftcor}, showing
rays through an image of the corona during eclipse.   The rays
intercept many different structures, and again the condition in 
Equation~(\pref{e:scales}) applies.  But this image mis\--represents the 
LOS confusion because the structures shown are already integrated
along the orthogonal LOS (in and out of the page).   
In 3D, the actual rays will intercept far fewer of these
structures than is suggested by this image.   {\em It is by no means
  clear that the LOS integration is worse in the corona than in the
  photosphere}, when it comes to trying to diagnose magnetic
  fields {\em of interest}.\footnote{When observing the photosphere on 
   larger scales, with a lower resolution (say $1^{\prime\prime}$; 725\,km),  the 
    magnetic flux tube structure shown is washed out.  The magnetic-field on
    the larger scales is still of interest, indeed most observations are made in
    this limit. However, the physical processes associated with
    flux tubes are not directly accessible to $1^{\prime\prime}$ resolution observations.}

%% Figure 
%

\subsection{Atomic Alignment}
\label{sec:align}

A proper interpretation of M1 emission-lines requires
knowledge of $\alignment$, in an inversion it must be solved for as part of
the solution for ${\bf S} \frt$ \citep{Judge2007}. 
The alignment 
comes from solutions to atomic sub-level population calculations.
Even in statistical-equilibrium, the equations are 
non-linear coupled multi-level systems requiring numerical
solution.  This presents a problem for inversions since 
this expands the
solution space to include the alignment itself, which becomes non linear in
the source parameters $
{\bf S} \frt  = \left
  \{\rho\frt,{\bf
  v\frt},
T\frt,{\bf B}\frt \right \}$.

To understand the non-linearities we can consider 
atomic models of increasing complexity.  First consider a 
two--level  atomic model  for a  $J=1  \rightarrow J_0=0$
transition  excited only by photospheric radiation, for which 
analytic solutions are available from, {\em e.g.}, 
\cite{Casini+Landi2008}.  Their  Equation 12.23
gives 
\begin{equation} \label{eqn:twol}
\sigma_0^2(1) = \frac{w}{2\sqrt{2}} \left (3 \cos^2 \vartheta_B -1
\right),\ \ {\rm (two\ level\ atom}, \ J=1 \rightarrow J_0=0)
\end{equation}
where $w$ measures the radiation anisotropy.  Here, $\vartheta_B$
(different from $\Theta_B$) measures the local angle between the
magnetic-field vector and solar gravity vector (central axis of the
radiation cone).  When the center-to-limb variation of the intensity
is zero, $w = \frac{1}{2} (1+\cos \vartheta_M) \cos \vartheta_M$
($\vartheta_M$ is the angle subtended by the solar radius at a point
[${\bf r}$] in the corona).  In this case the 
alignment is generated by anisotropic but rotationally symmetric
radiation in the transition itself.

The magnitude of alignment is reduced by processes
tending to populate sub-levels naturally, making 
$\frac{\pop{JM}}{\pop{J}} \rightarrow 1/(2J+1)$ and
so $ |\sigma^2_0(\alpha J)|\rightarrow 0$ 
in Equation
(\pref{eqn:alignpop}).  
Collisions with 
particles having isotropic distribution functions thus reduce
the magnitude of any existing alignment. Such collisions 
tend to leave the {\em angular dependence 
of existing alignment essentially unchanged}.   This 
result is demonstrated 
through the multi-level 
calculations for Fe {\sc XIII} by \cite{Judge2007}, where  
the alignment of the upper-levels ($3p^2\, ^3\!P_{J=2,1}$) 
of the 1074.7 and 1079.8\,nm lines of Fe~{\sc XIII} were 
found to factorize as
\begin{equation} \label{eqn:alignf0}
\sigma^2_0(\alpha_0 J) \approx\alig,
\end{equation}
to within $0.7\,\%$ and $3.2\,\%$ respectively.
The level-dependent term $k_J(T_{\rm e},n_{\rm e}, \vtm)$, an 
approximate 
generalization of the factor $\frac{w}{\sqrt{2}} $ in
Eq.~(\pref{eqn:twol}), is a
positive definite
factor depending only on local thermal conditions and the nature of
the disk irradiation through $\vtm$.  It is not linear in any of these
variables.   The factor 
$k_J(T_{\rm e},n_{\rm e}, \vtm)$ thus 
determines the magnitude of the alignment for any orientation of the
coronal magnetic field given by the other factor in variable 
$\vartheta_{\rm B}$.   
In Equations~(\pref{eqn:eps00}) and (\pref{eqn:eps31}) it enters
expressions for Stokes-$I$ and $V$ only as small first order corrections
that leave the signs of these terms unchanged. 

As a general rule the
magnitude of $k_J(T_{\rm e},n_{\rm e}, \vtm)$ is smaller for larger values of $J$,
since the number of sub states [$2J+1$] is larger. Thus
the 1079.8\,nm transition of Fe~{\sc XIII} ($J=2\rightarrow J=1$) has a
smaller
linear polarization than the 1074.7\,nm ($J=1\rightarrow 0$)
transition.   Transitions such as 1079.8\,nm with small 
$k_J(T_{\rm e},n_{\rm e}, \vtm)$ 
will therefore be useful since then the non-linear terms are commensurately
smaller in the Stokes-$I$ and $V$ parameters. 

For the $J=1$ level, Equation (\pref{eqn:twol}) represents an upper
limit to Equation (\pref{eqn:alignf0}), a limit which applies when
collisions are negligible ({\em e.g.,} $n_{\rm e} \rightarrow 0$).  The
alignment generated by anisotropic irradiation is reduced by sum of
all the collisions coupling the $J=1$ sub-levels to others in the 
26-level atom.  This behavior is expected in many other M1 lines of interest.

If the alignment can be shown to be zero, there is no linear
polarization and only the Stokes-$I$, $V$ profiles can be used to get a
``standard'' line-of-sight magnetogram for $B\cos\Theta$.  If it is
finite, it can take either sign because of the factor $(3\cos^2\vartheta_B - 1)$,  
and it leads directly to linear
polarization.  Observed minima in linear polarization, obtained for
example 
with the Coronal Multi-channel Polarimeter
[CoMP] \citep{Tomczyk+others2008}, often reflect the Van Vleck
condition $(3\cos^2\vartheta_B = 1)$,
 giving a direct indication
of part of the magnetic-field's geometry.  Passing across such
minima one finds a 90$^\circ$ change in direction of the linear
polarization vector as $(3\cos^2\vartheta_B - 1)$ and the alignment
changes sign, according to Equation~(\pref{eqn:epsi0}).  This is a
tell-tale sign of the Van Vleck effect even under the presence of
significant intergrations along the line-of-sight (LOS).  

%% Figure 
%
 \begin{figure} 
\vskip 10pt
 \centerline{
\includegraphics[width=\textwidth]{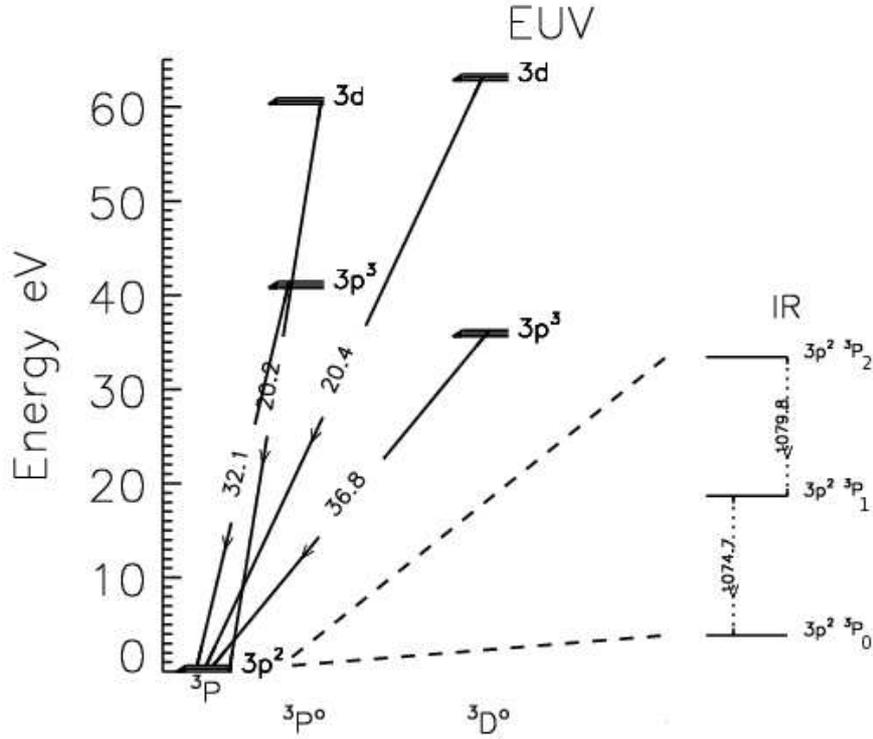}
}
 \caption{Term diagram for Fe {\sc XIII} showing the strongest E1
   transitions of each multiplet,  and 
the M1 lines among the levels of the ground term.   The 1074.7\,nm
line has upper-level $J=1$, lower $J=0$.  
Each configuration
shown has EUV line ratios sensitive to density and 
photospheric radiation field as a result of the competition for sub-level
populations in the ground term.  
}\label{fig:fe13_td}
\end{figure}

With these arguments 
we can summarize the role of the alignment as follows 
({\em e.g.} \citealp{Judge2007}): 
\begin{enumerate}
\item {} The magnetic-field {\em azimuth} has the well\--known $90^
\circ$ ambiguity, {\em unless} 
the sign of $\alignment$ can be determined, in which case there
remains a 
180$^\circ$ ambiguity.
\item{} The magnitude and sign of the alignment $\alignment$ affects all
  four Stokes parameters.   
\item{} Measurements of electron\--density\--sensitive lines at 
IR and EUV wavelengths will help determine $|\alignment|$ and should
be included as part of the vector of observables ${\bf I}$. 
\item{} Measurements of M1 lines from $J>1$ levels ({\em e.g.} Fe~{\sc XI} 782.9\,nm,
  Fe~{\sc XIII} 1079.8\,nm) with their smaller alignment $|\alignment|$, 
will make inversions more  linear. In comparison with 
strongly aligned transitions (Fe {\sc XIII} 1074.7\,nm for example), such transitions
 have  smaller 
$|\alignment|$ non-linear factors for $I$ and $V$ in Equations 
(\pref{eqn:eps00}) and (\pref{eqn:eps31}). 
\end{enumerate}

\subsection{Selection of Lines for Inversion}

 \cite{Judge2007} has examined how the alignment 
might be constrained -- even determined -- from observations, in the simplest case where a single
point dominates all emission from an M1 coronal line.  For a given set
of such measurements [{\bf I}],  he has shown that
there are generally multiple roots to the governing Equations for the
atomic alignment.  The solutions correspond to different scattering
geometries that are compatible with data (see his Table 2).  
Even in principle there is
no unique solution.

%% Table
%
 \begin{table}
\caption{An example of a set of lines in Fe~{\sc XIII} for magnetic inversions}\label{tbl:1}
\begin{tabular}{llll}
%\hline
$\lambda$   & Type & Data & Transition and Comments \\
            \ [nm]    &         & needed &\\   
%\hline
1074.7 & M1 & IQUV & ${3p^2} \, ^3\!P_1 - 3p^2 \, ^3\!P_0$,\  large $|\alignment|$\\
1079.8 & M1 & IQUV & ${3p^2} \, ^3\!P_2 - 3p^2 \, ^3\!P_1$,\  
small $|\alignment|$\\
35.97 & E1  & I & ${3s3p^3} \, ^3\!D^0_{1,2} - 3p^2 \, ^3\!P_1$, blend
of two lines \\
34.82 & E1  & I & ${3s3p^3} \, ^3\!D^0_1 - 3p^2 \, ^3\!P_0$ \\
20.38 & E1 & I & $3p3d\, ^3\!D^0_3 - 3p^2\, ^3\!P_2$\\
20.20 & E1  & I & $3p3d\, ^3\!P^0_1 - 3p^2\, ^3\!P_0$\\
%\hline
\end{tabular}
\end{table}

However, Judge considered a dataset consisting of just one M1 line. 
From section \pref{sec:align}, it is clear that the inversion problem
will benefit from more data that can restrict the range of 
thermal conditions that, at each point in the corona, are compatible
with data.   In effect this will limit the level-dependent factor 
[$k_J(T_{\rm e},n_{\rm e}, \vtm)$] in $\alignment$.  

\begin{figure} 
 \centerline{
\includegraphics[width=0.5\textwidth,clip=0]{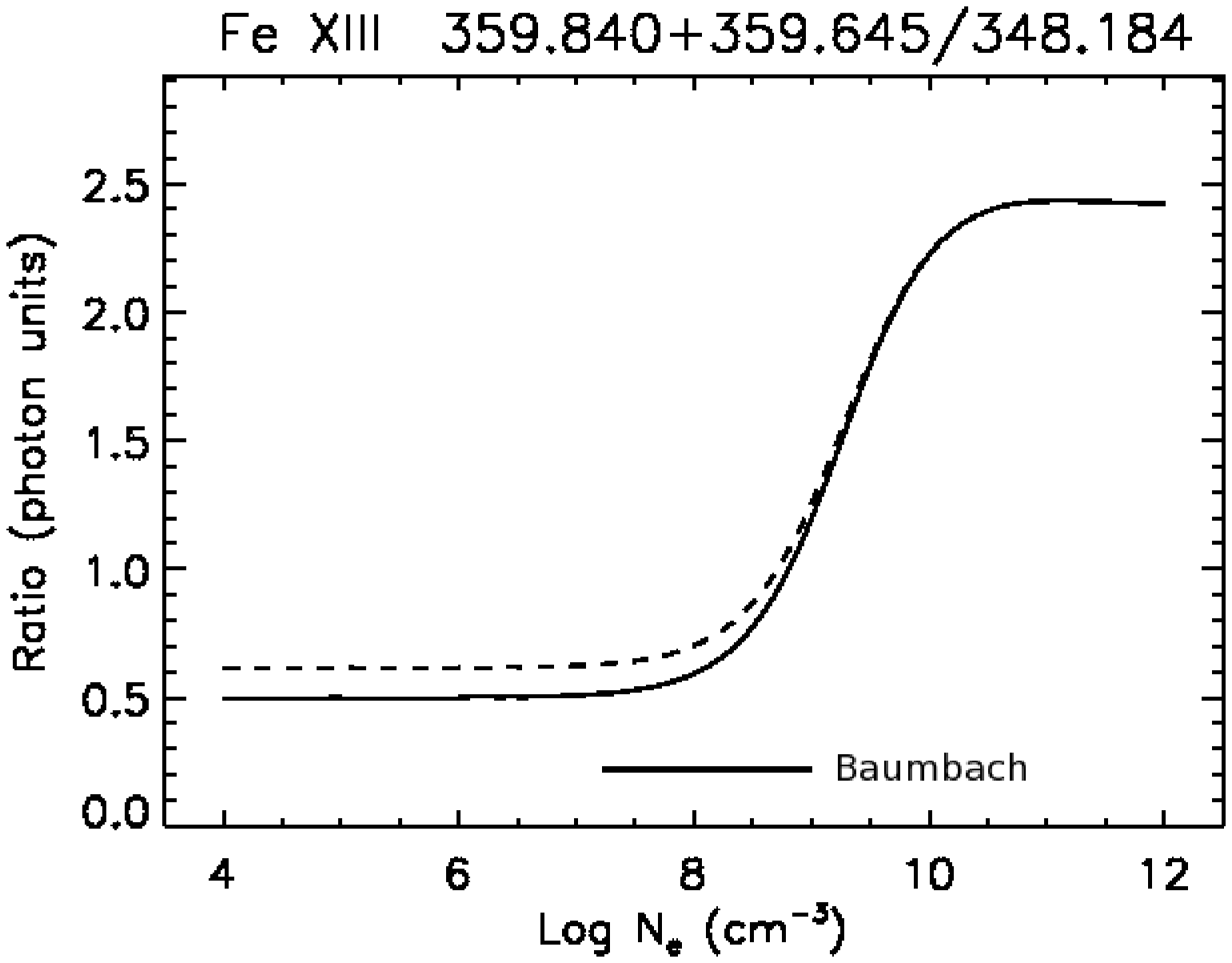}
\includegraphics[width=0.5\textwidth,clip=0]{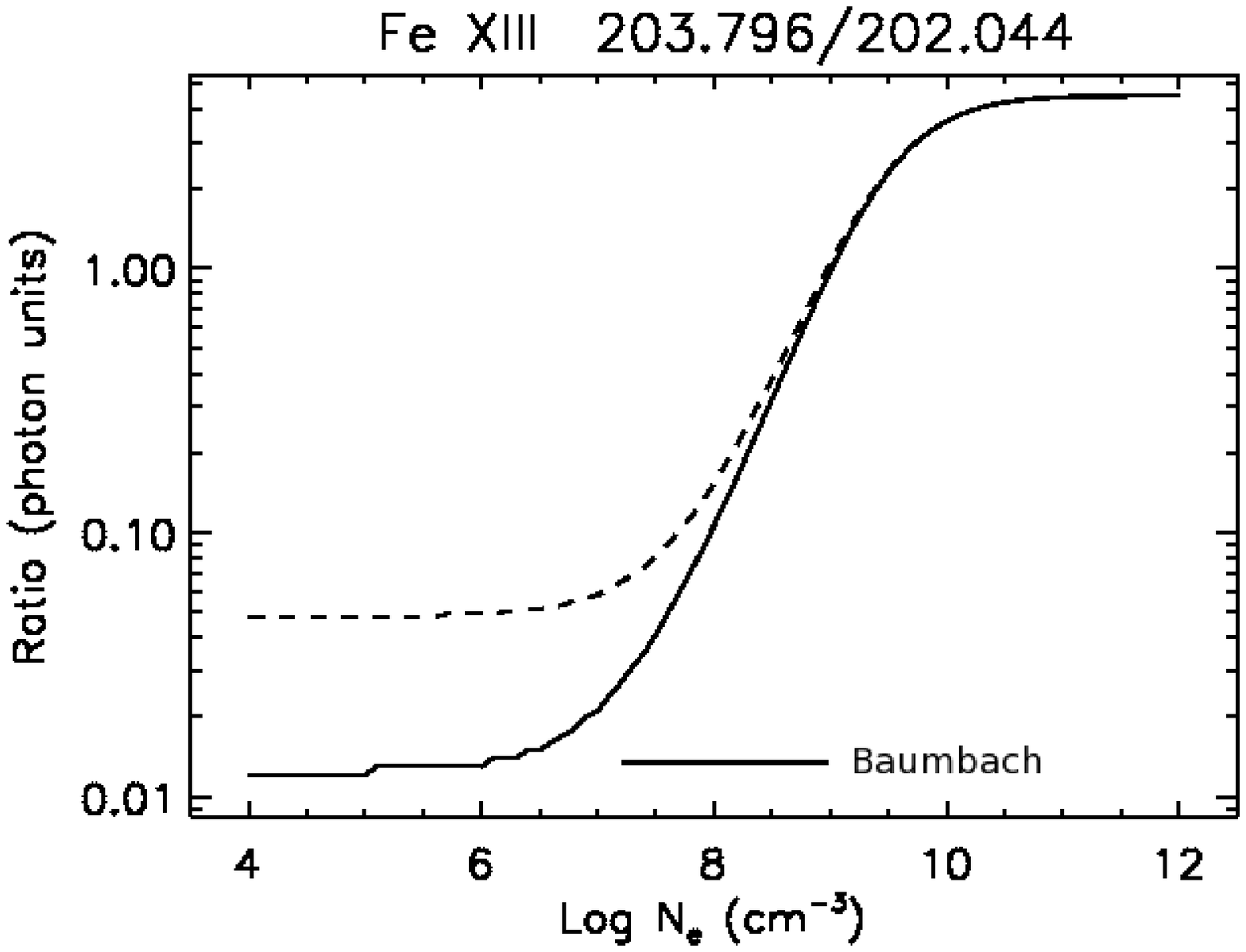}
}
 \caption{Typical density-sensitive line intensity 
ratios computed 
for Fe {\sc XIII}.  
Left panel: intensity ratios from 
a blend of two lines near 35.97\,nm to another line, all within
the $3s3p^3\, ^3\!D^o - 3s^23p^2\, ^3\!P$ 
 multiplet. Right panel: a ratio of two lines within the 
$3s^23p3d\, ^3\!D^o - 3s^23p^2\, ^3\!P$ 
multiplets. 
Dashed lines  
include radiative excitation, solid lines do not.  Note that the
ratios are sensitive in a density regime of interest [$n_{\rm e} \approx 10^8 $
cm$^{-3}$].   Note that the wavelengths are in \AA{} units not nm in
the figure, and that atomic alignment is ignored in these
calculations.  
The line marked ``Baumbach'' shows typical variations in
electron density 1.005 to about 1.4 ${\rm R}_\odot$ as given by Baumbach in
\cite{Allen1973}.  
}\label{fig:fe13dens}
\end{figure}

Both M1 and E1 EUV spectral lines contain temperature- and density-
sensitive lines which can be used to help determine $k_J(T_{\rm e},n_{\rm e},
\vtm)$, thereby helping resolve ambiguities inherent in using single
M1 lines.  Thus {\em the data to be inverted should be expanded to include
a variety of lines}.  Let us focus on Fe {\sc XIII} as a concrete example.
A term diagram is shown in Figure \pref{fig:fe13_td}.  
Fe {\sc XIII} (Si - like) has a density-sensitive pair of M1 lines (1074.7
and 1079.8\,nm) as well as various pairs in the EUV.  These arise
mainly because of the competing roles of radiative excitation, de-excitation,
and collisions in determining the (sub) level populations among the
ground $^3\!P_{J=0,1,2}$ term.   Figure~\pref{fig:fe13dens} shows ratios of
EUV lines for Fe {\sc XIII} that are sensitive to the radiation field and
electron density, together with the range of densities expected in the
low corona from Section 84 of \cite{Allen1973}.  
Table 
\pref{tbl:1}
lists 
various transitions that might be observed and put into the ``vector
of observations'' [${\bf I}$] for inversion.   
Joint CoMP and EUV  measurements with the EIS instrument on 
the {\em Hinode} spacecraft have already 
been made in  August and November 2012, including 
the Fe {\sc XIII} lines of 1074.7, 10798., 20.38 and 20.20\,nm.
Since CoMP observes almost daily, there will be other 
observations where yet more Fe {\sc XIII} lines are available for analysis.  
%We plan to analyze these data with the CoMP group at HAO. 

%% Figure 
%

\begin{figure} 
 \centerline{
\includegraphics[width=0.9\textwidth,clip=0]{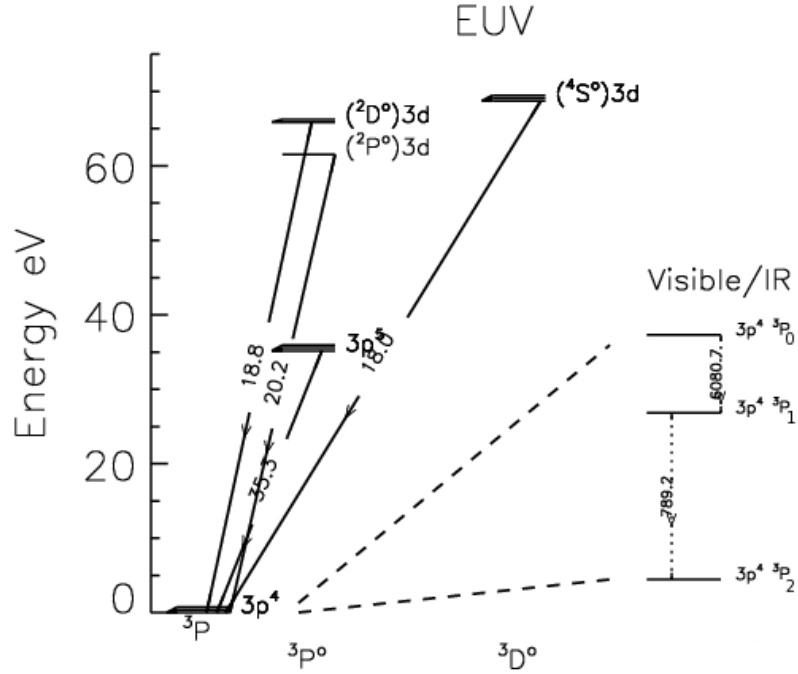}
}
 \caption{Term diagram for S-like Fe {\sc XI}.  Unlike Si-like 
Fe {\sc XIII}, the ground level 
has $J=2$, the 789.2\,nm transition occurs between the 
upper $J=1$ level and the ground level.   The $J=0\rightarrow J=1$ is
at the anomalously large wavelength near 6082\,nm because the Fe {\sc XI}
ion's levels are close together as $jj$ coupling competes with $LS$
coupling (see Judge 1998).  
}\label{fig:fe11_td}
\end{figure}

Other suitable ions 
(from the bright lines computed by \citealp{Judge1998})
include S-like Fe {\sc XI} with an M1 line near 789.2\,nm, B-like Mg {\sc VIII} (3028\,nm), C-like Si {\sc IX} (3934\,nm), and of course
the red and green coronal lines (Cl-like Fe {\sc X} and Al-like Fe {\sc
  XIV}
respectively).  There are various pros/cons with the selection of
lines.  For example Fe {\sc XI} 789.2\,nm shows remarkable structure in eclipse
images \citep{Habbal+others2011}, it lies in the near infrared so has
low stray light and reasonable sensitivity to the Zeeman effect.
It is formed at lower temperatures than Fe {\sc XIII} and hence may be 
useful in cooler regions of the corona, say over coronal holes. 
As noted, this line is expected to have a small atomic alignment 
so that although the linear polarization will be small
in 789.2\,nm, so will the alignment corrections to the emission
coefficients for Stokes-$I$ and
$V$.   It should therefore be considered as a {\em prime target} for
future observations.   A term diagram for 
Fe {\sc XI} is shown in 
Figure \pref{fig:fe11_td}.  Cases can also be made for the other
strong lines of various ions discussed by \cite{Judge1998}.

%% Figure 
%

\section{Tomographic-Inversions}

Slicing through the volume containing magnetic-fields by observing
lines of sight at different angles opens up the possibility of full 3D
vector-field recovery.  Some studies of vector tomography have
been made by\footnote{Note that their studies
  are naturally in the strong-field limit of the Hanle effect,
  although they refer (inaccurately) to ``the Hanle effect''. }
\cite{Kramar+Inhester+Solanki2006,Kramar+Inhester2007}.  These are
preliminary in that they explore either $I,V$ or $I,Q,U$, not the full
Stokes vector.    Further, just one theoretical emission-line was
``inverted''
so that the observed data contain limited information on the alignment 
$\alignment$.     They conclude however that 
\begin{quote}
  ``We are confident that this data set is also sufficient to yield a
  realistic coronal magnetic-field model. This, however, has to be
  verified in future [numerical] experiments.''
\end{quote}
Their method attempts to handle the existence of null spaces in the
inversion by standard techniques of adding a ``regularization'' 
parameter.   Thus far they have investigated the minimization 
of the functional
\begin{equation}
  \label{eq:functional}
   L({\bf B}) = \mu ({\bf I_{\rm OBS}-I_{\rm SIM}}) \cdot ( {\bf I_{\rm OBS}-I_{\rm SIM}} )
   + \int {\rm div^2 }\, {\bf B}\, {\rm d}^3V
\end{equation}
where the integral is over the coronal volume.  Minimization of $
L({\bf B}) $ simply forces the selection of a 3D magnetic-field to
minimize the differences between observed ${\bf I}_{\rm OBS}$ and computed
${\bf I}_{\rm SIM}$ intensities and polarized Stokes parameters, subject
to the additional constraint depending on $\mu$.  $\mu$ is a parameter
that determines how much of the solution is determined by the data
($\mu$ large) and by the physically imposed divergence constraint
($\mu$ small).  (Note that $\mu$ should include the 
estimates of the observed uncertainties for each component of the 
vector of observables too).

The divergence constraint alone means that the space of curl-free
vector-fields is a null space: potential field components along the
LOS are invisible to Stokes-$IV$.  They speculate that by adding the
force-free constraint into the regularization (as $\int |{\bf J \times
  B}|^2{\rm d}^3V$), this null space might be eliminated.

It should be remembered that such inversions rely on 
stereoscopic observations of coronal M1 lines
(not currently possible) or on the
assumption that the corona is a solidly rotating body, observed 
from the Earth over periods of at least a day.    

If we combine our understanding from 
section \pref{sec:inversion} with tomography, we see that
with a general forward modeling code such as that 
written by 
\cite{Judge+Casini2001}, we can 
in principle  invert a vector of
observations including M1 lines with large and small alignment
factors and selected E1 lines, to obtain the desired solutions for 
${\bf B}\frt$.    Key to this effort will be the regular detection of the
Stokes-$V$ parameters of M1 lines, something that has not yet been
achieved owing to the small apertures of coronagraphs currently used. 
Unpublished work by Judge using the prototype 
CoMP instrument ($d=20$\,cm) 
acquired in February 2012
gives an upper limit of 0.15\,\% for the maximum ratio of $V/I$ in
1079.8\,nm.    In 70 minute integrations and a
low  (20$^{\prime\prime}$) spatial resolution,  \cite{Lin+Kuhn+Coulter2004} achieved 
a sensitivity below 0.01\%, leading to a Stokes-$V$ amplitude over an active
region of about $0.0001I$, with a 0.46\,m diameter coronagraph.  

Clearly, bigger telescopes are needed at excellent sites for this kind
of work to succeed.  The  COronal Solar Magnetism Observatory [COSMO] offers one possible solution.

\section{Discussion}

The tomographic-inversion scheme outlined above is the only way to
invert formally data vectors to recover the coronal ${\bf B}\frt$. The
scheme relies on solar rotation and assuming the coronal structures
are stationary over periods of a day or longer, or on the future
availability of stereoscopic measurements both from earth and from a
spacecraft (like the Solar TErrestrial RElations Observatory [STEREO]) at a significant elongation
from the earth.  The latter possibility has yet to be discussed at all
and so is 
decades away.   The former 
is naturally limited, but should be pursued once regular observations
of the weak Stokes-$V$ signal are available.  The CoMP instrument 
is a prototype for larger instruments which should achieve this
goal ({\em e.g.}, the Advanced Technology Solar Telescope [ATST], COSMO).

\subsection{Local Analyses of Coronal Loops}

It seems prudent also to relax our goal of reconstructing ${\bf B}\frt$ 
via tomography and look to other ways that we
might make progress in this area.  One possibility is to assume 
that we can identify a single plasma loop in an M1 transition,
as routinely done for EUV or X-ray data. In such a case the source
vector [${\bf S}$] only has contributions predominantly from lines of
sight that intersect the loop.  Also let us assume that observations
from another viewpoint (EUV data from STEREO for example) are
available that fix the heliocentric coordinates of the plasma loop.  This
additional information enables us to diagnose magnetic-fields beyond
what is possible from an isolated measurement of the Stokes profiles
of a single point \citep{Judge2007}.  However, as for EUV lines, 
no useful information outside the plasma loop volume is available.
Nevertheless this should be pursued.

\subsection{Direct Synthesis {\em vs} Observations}

Another avenue to explore adding information to the data is to assume that we
know more about the current\--carrying structures that we are looking for.  Thus,
by building synthetic maps of M1 lines from models of the magnetic
field and coronal plasma, and comparing them directly
with observations, one can hope to extract meaningful information.  It
may be possible to argue that the data are inconsistent with a class
of model (``i''), whereas another class (``ni'') is not inconsistent.  Science
advances often by identifying models of class (i), those of class (ni)
being acceptable subject to further investigation.  This will be a
fruitful approach; already some initial comparisons reveal models of type
(ni) (\citealp{Rachmeler+others2012}, in this volume) but as of yet we
are not aware interesting cases in class (i). There are obvious cases where
potential fields, extrapolated from the lower atmosphere fall 
into class (i),   but this
finding serves merely to show that some free magnetic energy 
appears necessary to describe coronal structures.  This is something 
we have known for decades through other arguments
({\em e.g.} \citealp{Gold+Hoyle1960}). 

These are early days though.
The main issue with this approach is that 
\begin{quote}{\em There are more things in heaven and earth, Horatio, 
than are dreamt of in your philosophy.} -- Hamlet
\end{quote}

\subsection{Closing thoughts on the ``line-of-sight problem''} 

Consider the idea that in highly conducting plasma, one can
trace magnetic-fields by looking at morphology of plasma loops.  This
was a motivation for the Transition Region and Coronal Explorer
[TRACE] mission  (hence its name) and it has
yielded many such morphological analysis of ``coronal magnetism'',
including seismology (one nice example is that of 
\citealp{Aschwanden+others1999}).  Apparently
the LOS issues do not present special challenges in these analyses of
coronal-intensity measurements. One might argue that these are seen 
against the dark solar disk (any  
EUV continuum emission from the low temperature 
photosphere/low chromosphere is very dark), whereas
the M1 coronal lines must be observed
above the limb against a dark background.  But even in this
case, isolated bright plasma loops organized into an active region 
offer no greater path lengths for
integration than observations on the disk.  Indeed, the discovery 
of MHD wave modes in the M1 Fe~{\sc XIII} 1074.7\,nm line 
\cite{Tomczyk+others2007} indicates that, just as for EUV work, 
line of sight confusion is not an overwhelming problem. 

We conclude that, {\em as in all remotely sensed magnetic data}, 
line of sight issues are important but not intractable.  
Often, using M1 lines we will be 
interested in the coronal magnetic-fields above active regions. These 
present themselves as bright isolated plasma loops in M1
coronal lines just as the EUV and X ray lines do \citep{Bray+others1991},
dominating the contributions to the Stokes vectors along the line of
sight.

\section{Conclusions}\label{s:Conclusions} 

Scientific skepticism is healthy, and we certainly need to be
skeptical of interpretations of all remotely sensed data of an object
like the Sun.  We have shown that the optically thin forbidden coronal
lines suffer from the same kinds of interpretational problems as do
other diagnostics of solar magnetism.  We have suggested several ways
to augment the data of isolated points in the corona - for which we
have vast null spaces of unexplorable parameters - using tomography
and traditional ideas concerning the smoothness and continuity of
magnetic-fields in coronal structures, applied universally to EUV and
X-ray intensity data.  

It will be interesting to see how a full vector inversion including
lines sensitive to thermodynamic parameters -- both visible/IR M1 lines
and EUV lines -- will serve to further constrain tomographic-inversions.
Certain schemes (especially ``direct [matrix] 
inversions'') can be very fast, but these
require linear Equations which is manifestly not the case (see
the Equations above).   It is, however, possible that the non-linearities
introduced by the alignment into these equations can be treated to
some degree by a formal (Newton-Raphson- style) linearization scheme.
This seems promising given that we have lines with quite different 
alignment factors (1074.7 vs. 1079.8 or 798.2\,nm) and 
thus different non-linear amplitudes, but this is an area that 
remains to be explored.

Several ways forward are reviewed while we
await the arrival of high-sensitivity ($\lesssim 10^{-4}$)
polarization data from telescopes (ATST, COSMO) needed for tomographic
inversions that can recover the vector-field throughout volumes of the
corona.

%%% %%%%%%%%%%%%%%%%%%%%%%%%%%%%%%%%%%%%%%%%%%%%%%%%%%%%%%%%%%%
%% Bibliography
%
% Using BibTeX
%

%\fullreferences
%\aareferences
%\apjreferences

\end{article} 
\end{document}